\documentclass[aps,pra,floatfix,preprint,nofootinbib,superscriptaddress,showpacs]{revtex4}
\usepackage{placeins}
\usepackage{latexsym}
\usepackage{amssymb}
\usepackage{amsmath}
\usepackage{graphicx}
\usepackage{hyperref}
\usepackage{cases}
\usepackage{longtable}
\newcommand{\fl}[1]{\hspace*{-\mathindent}\parbox{.97\linewidth}{\raggedright{$#1$}}}
\newcommand{\lo}[1]{\hspace*{-\mathindent}\parbox{.97\linewidth}{\qquad\raggedright{$#1$}}}

\begin{document}

\title{Alternative Mathematical Technique to Determine LS Spectral Terms}

\author{Renjun Xu}
\email[E-mail: ]{xu5n@yahoo.com}
 \affiliation{National Laboratory
of Super Hard Materials, Jilin University, Changchun 130012, P.R.
China}
 \affiliation{College of Physics, Jilin University, Changchun
130023, P. R. China}
\author{Zhenwen Dai}
\email[E-mail: ]{jzk@mail.jlu.edu.cn}
 \affiliation{College of Physics, Jilin University, Changchun
130023, P. R. China}
 \affiliation{Key Lab of Coherent Light, Atomic and Molecular Spectroscopy, Ministry of Education, Jilin University, Changchun
130023, P.R. China}


\begin{abstract}
We presented an alternative computational method for determining
the permitted LS spectral terms arising from $l^N$ electronic
configurations. This method makes the direct calculation of LS
terms possible. Using only basic algebra, we derived our theory
from LS-coupling scheme and Pauli exclusion principle. As an
application, we have performed the most complete set of
calculations to date of the spectral terms arising from $l^N$
electronic configurations, and the representative results were
shown.\footnote{The table of LS terms for $l^N(l=0$--$5)$
configurations is too long and over two pages, so we only present
part of the result in this article.}~ As another application on
deducing LS-coupling rules, for two equivalent electrons, we
deduced the famous Even Rule; for three equivalent electrons, we
derived a new simple rule.
\end{abstract}

 \maketitle

\section{INTRODUCTION}

In the atomic and nuclear shell model, a basic but often laborious
and complicated problem is to calculate the spectral terms arising
from many-particle configurations. For a N-particle occupied
subshell $l^N$, currently we often use computational methods based
on the unitary-group representation theory, which have been
developed by Gelfand et al.~\cite{gt}, M. Moshinsky et
al.~\cite{mm}, Biedenharn et al.~\cite{bied}, Judd~\cite{judd},
Louck et al.~\cite{louck}, Drake et al.~\cite{drake}, Harter et
al.~\cite{hart}, Paldus~\cite{pal}, Braunschweig et
al.~\cite{bh78}, Kent et al.~\cite{ks81} and others, extending
thus the classical works of Weyl~\cite{weyl}, Racah~\cite{racah},
S. Meshkov~\cite{sm53}, Cartan, Casimir, Killing, and others. For
efforts of all these works, can we have the current calculation
method much more simplified and less steps needed than ever
before. However, when many electrons with higher orbital angular
momentum are involved in one subshell, the calculation process
using this theoretical method still is a challenging work. The
current feasible methods usually take several steps of
simplification, such as firstly using Branching Rules for
reduction~\cite{bh78}, and then using LL-coupling scheme.
Nevertheless, we still have to work hard to calculate a big table
of the LS terms corresponding to Young Patterns of one column (the
situation is similar to use Gelfand basis
set~\cite{hart,pal,pal74}), and many LL-couplings. Often this is a
difficult and complicated job.

In this paper, we present an alternative mathematical technique
for direct determination of spectral terms for $l^N$
configurations. The new theory consists of a main formula
[equation~(\ref{main})] and four complete sub-formulas
[equations~(\ref{sub1})-(\ref{sub4})], all of which are common
algebra expressions. The basis of this method does not require any
knowledge of group theory or other senior mathematics.

The organization of this paper is as follows: the five basic
formulas and some notations are introduced in Section\;II (the
derivations of those formulas are presented in Appendix). The
specific calculation procedure is shown in Section\;III. In
Section\;VI, as some applications using this alternative theory,
we presented permitted spectral terms for several $l^N$
configurations; then deduced naturally the well-known Even Rule
for two electrons; and for three electrons, we derived a new
compact rule. Finally, conclusions are drawn in Section\;V.

\section{THEORETICAL OUTLINE AND NOTATIONS}

\subsection{Notations}

In the following, we denote by $X(N, l, S', L)$ the number of
spectral terms with total orbital angular quantum number $L$ and
total spin quantum number $S'/2$ arising from $l^N$ electronic
configurations. (To calculate and express more concisely, we
doubled the total spin quantum number $S$ and the spin magnetic
quantum number $M_S$ here, which are correspondingly denoted by
$S'$ and $M_S'$. Hence, all discussions in the following are based
on integers.) When the function $X(N,l,S',L)=0$, it means that
there is no spectral terms with total orbital angular quantum
number $L$ and spin quantum number $S'/2$. We denote by $A(N, l,
l_b, M_S', M_L)$ the number of LS terms having allowable orbital
magnetic quantum number $M_L$ and spin magnetic quantum number
$M_S'/2$, arising from $l^N$ electronic configurations. $l_b$ is
defined as the largest allowable orbital magnetic quantum number
$(m_{l_i})_{\max}$ in one class. Its initial value equals $l$
according to equation~(\ref{main}).

\subsection{The Complete Basic Formulas}

The main formula to calculate the number of LS terms in $l^N$
electronic configurations is given below,
\begin{eqnarray}
X(N, l, S', L)=A(N, l, l, S', L)- A(N, l, l, S',
L+1)\nonumber \\
{}+ A(N, l, l, S'+ 2, L+1)- A(N, l, l, S'+2, L),
 \label{main}
\end{eqnarray}
where the value of function $\mathbf{A}$ is based on the following
four sub-formulas
\begin{widetext}
\begin{eqnarray}
\fl{{\rm Case\ 1:\ } {M_S'=1},\ \vert M_L \vert \leq l,\
{\rm and \ } {N=1}} \nonumber \\
\lo{A(1,l,l_b,1,M_L)=1} \label{sub1} \\
\fl{{\rm Case\ 2:\ } {\begin{array}{ll}
\{M_S'\}=\{2-N,4-N,\ldots,N-2\},\\
\vert M_L \vert \leq f(\frac{N-M_S'}{2}-1)+f(\frac{N+M_S'}{2}-1),\
{\rm and \ }1<N \leq 2l+1
\end{array}}} \nonumber\\
\lo{A(N,l,l,M_S',M_L)=~\hspace{-1.5em}\sum\limits_{M_{L_-}=\left\{-f(\frac{N-M_S'}{2}-1),\;M_L-f(\frac{N+M_S'}{2}-1)\right\}_{\max
}}^{\left\{f(\frac{N-M_S'}{2}-1),\;M_L+f(\frac{N+M_S'}{2}-1)\right\}_{\min}}
\hspace{-1.65em}\left\{
A\left(\frac{N-M_S'}{2},l,l,\frac{N-M_S'}{2},M_{L_-}\right)
\right. }
\nonumber\\
\lo{\hphantom{A(N,l,l,M_S',M_L)=}\qquad\times \left. A\left(
\frac{N+M_S'}{2},l,l,\frac{N+M_S'}{2},M_L-M_{L_-}\right) \right\}}
\label{sub2}\\
\fl{{\rm Case\ 3:\ } M_S'=N,\ \vert M_L \vert \leqslant f(N-1),\ {\rm and \ } 1<N \leq 2l+1} \nonumber\\
\lo{A(N,l,l_b,N,M_L)=~\hspace{-1em}\sum\limits_{M_{L_I}=\lfloor
\frac{M_L-1}{N}+\frac{N+1}{2}
\rfloor}^{\{l_b,\;M_L+f(N-2)\}_{\min}}
A(N-1,l,M_{L_I}-1,N-1,M_L-M_{L_I})} \label{sub3}\\
\fl{{\rm Case\ 4:\ } {\rm other~cases~just~do~not~exist, \ therefore}}\nonumber \\
\lo{A(N,l,l_b,M_S',M_L)=0} \label{sub4}
\end{eqnarray}
\end{widetext}
where the floor function $\lfloor x \rfloor$ presented in this
paper denotes the greatest integer not exceeding $x$, and
\begin{equation*}
f(n)=\left\{ {\begin{array}{*{20}l}
{\sum\limits_{m=0}^n(l-m)} & {for~n \geq 0}  \\
{0} & {for~n<0}  \\
\end{array}}\right.
\end{equation*}
The derivations of equations~(\ref{main})-(\ref{sub3}) are
presented in detail in Appendix.

\section{THE SPECIFIC PROCEDURE}

\begin{figure*}[!htb]
\centering
\includegraphics[width=\textwidth]{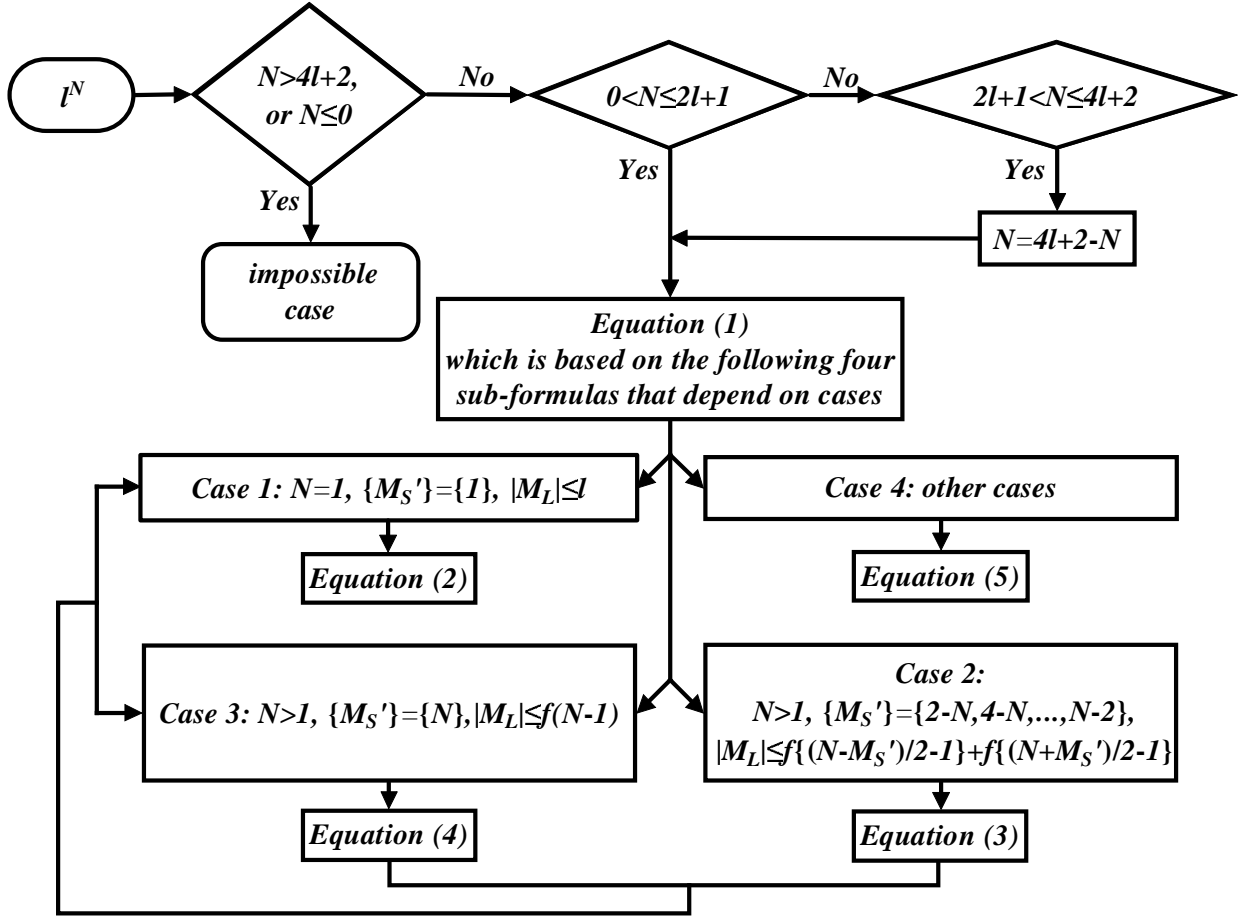}
\caption{Specific procedure to determine LS terms arising from
$l^N$.} \label{figI}
\end{figure*}

A concrete procedure to determine the LS spectral terms arising
from $l^N$ electronic configurations is given in
Figure~\ref{figI}.
For $l^N$ electronic configurations, if $N$ is larger than
$(2l+1)$ and less than $(4l+2)$, it is equivalent to the case of
$(4l+2-N)$ electrons; Else if $N$ is not larger than $2l+1$, the
total spin quantum number $S$ could be $\{\frac{N}{2}-\lfloor
\frac{N}{2} \rfloor,\frac{N}{2}+1-\lfloor \frac{N}{2}
\rfloor,\ldots,\frac{N}{2}\}$ [equations~(\ref{range:S'},
\ref{range:S2})], and the total orbital angular quantum number $L$
could be
$\{0,1,\ldots,f(\frac{N-M_S'}{2}-1)+f(\frac{N+M_S'}{2}-1)\}$
[equation~(\ref{range:L})]. The number of LS terms with total
orbital angular quantum number $L$ and total spin quantum number
$S$ $(S=S'/2)$ is calculated by function $X(N,l,S',L)$. Based on
equation~(\ref{main}), then the main task is to calculate the
function $A(N,l,l_b,M_S',M_L)$. Due to the value of three
parameters $N$, $M_S'$, and $M_L$ in function $\mathbf{A}$, there
are four cases. If it is in the condition of case 2 or case 3, we
can calculate the function $\mathbf{A}$ based on the
equation~(\ref{sub2}) or equation~(\ref{sub3}), both of which
could come down to case 1 or case 3. Finally, we will get the
eigenvalue of function $\mathbf{X}$. If the function $\mathbf{X}$
vanishes, it means that there is no corresponding LS terms.

\section{EXAMPLES and APPLICATIONS}

\subsection{Permitted LS terms of $l^N$ subshell}

Based on the flow chart shown in Figure~\ref{figI}, we have
written a computer program in C language. For the length limit of
this article, we only presented in Table~\ref{tab:1} the LS terms
for $g^{9}$ and $h^{11}$ electronic configurations. (The terms for
$s^N,p^N,d^N$, and $f^N$ can be found in Robert D. Cowan's
textbook~\cite[p.~110]{cowan}.) As far as we know, LS spectral
terms of $g^N$ and $h^N$ given here are reported for the first
time in literature.

The notation of the spectral terms given below is proposed by
Russell \cite{russell} and Saunders \cite{russell25} and now has
been widely used.

${\begin{array}{*{26}l}
 {L = } & 0 & 1 & 2 & 3 & 4 & 5 & 6 & 7 & 8 & 9 & {10} & {11} & {12}\\
  & S & P & D & F & G & H & I & K & L & M & N & O & Q\\
 {L = } & {13} & {14} & {15} & {16}& {17} & {18} & {19} & {20} & 21 & 22 & 23 & {\cdots}\\
  & R & T & U & V & W & X & Y & Z & 21 & 22 & 23 & {\cdots}\\
\end{array}}$\\
When the orbital quantum number $L$ is higher than $20$, it is
denoted by its value. Owing to the length of the table, a compact
format~\cite{gww} of terms is given here: $^A(L_{k_1}L_{k_2 }^{'}
\ldots )$, in which the superscript $A$ indicates the multiplicity
of all terms included in the parentheses, and the subscripts
$k_1$, $k_2$ indicate the number of terms, for example $^2G_6$
means that there are six $^2G$ terms.

\begingroup
\setlength{\LTcapwidth}{\textwidth}
\begin{longtable*}{@{}c p{0.825\linewidth}}
\caption[Permitted LS terms for selected $l^N$
configurations]{\label{tab:1} Permitted LS terms for selected
$l^N$
configurations.}\\[-1em]
\hline \hline \multicolumn{1}{c}{Configurations} &
\multicolumn{1}{c}{LS spectral
terms}\\
\hline
\endfirsthead
\caption{Permitted LS terms for selected $l^N$
configurations (continued).}\\[-1em]
\hline \hline \multicolumn{1}{c}{Configurations} &
\multicolumn{1}{c}{LS spectral
terms}\\
\hline
\endhead
\hline \hline \endfoot \hline \hline \endlastfoot
$g^9 $&$ ^2(S_8\!$ $P_{19}\!$ $D_{35}\!$ $F_{40}\!$ $G_{52}\!$
$H_{54}\!$ $I_{56}\!$ $K_{53}\!$ $L_{53}\!$ $M_{44}\!$ $N_{40}\!$
$O_{32}\!$ $Q_{26}\!$ $R_{19}\!$ $T_{15}\!$ $U_9\!$ $V_7\!$
$W_4\!$ $X_2\!$ $YZ)$ $^4(S_6\!$ $P_{16}\!$ $D_{24}\!$ $F_{34}\!$
$G_{38}\!$ $H_{40}\!$ $I_{42}$ $K_{39}\!$ $L_{35}\!$ $M_{32}\!$
$N_{26}\!$ $O_{20}\!$ $Q_{16}\!$ $R_{11}$ $T_7\!$ $U_5\!$ $V_3\!$
$W\!\!$~$X)\ $
$^6(S_3P_3D_9F_8G_{12}H_{10}I_{12}K_9L_9M_6N_6O_3Q_3RT)\ $
$^8(PDFGHIKL)\ $ $^{10}(S)$\\
$h^{11} $&$ ^2(S_{36}\!$ $P_{107}\!$ $D_{173}\!$ $F_{233}\!$
$G_{283}\!$ $H_{325}\!$ $I_{353}\!$ $K_{370}\!$ $L_{376}\!$
$M_{371}\!$ $N_{357}\!$ $O_{335}\!$ $Q_{307}\!$ $R_{275}\!$
$T_{241}\!$ $U_{207}\!$ $V_{173}\!$ $W_{142}\!$ $X_{114}\!$
$Y_{88}\!$ $Z_{68}\!$ $21_{50}\!$ $22_{36}\!$ $23_{25}\!$
$24_{17}\!$ $25_{11}\!$ $26_7\!$ $27_4$ $28_2\!$ $29$ $30)\ $
$^4(S_{37}\!$ $P_{89}\!$ $D_{157}\!$ $F_{199}$ $G_{253}\!$
$H_{277}\!$ $I_{309}\!$ $K_{313}\!$ $L_{323}\!$ $M_{308}$
$N_{300}\!$ $O_{271}\!$ $Q_{251}\!$ $R_{216}\!$ $T_{190}\!$
$U_{155}$ $V_{131}\!$ $W_{101}\!$ $X_{81}\!$ $Y_{59}\!$ $Z_{45}\!$
$21_{30}\!$ $22_{22}$ $23_{13}\!$ $24_9\!$ $25_5\!$ $26_3\!$ $27$
$28)\ $ $^6(S_{12}\!$ $P_{35}$ $D_{55}\!$ $F_{76}\!$ $G_{90}\!$
$H_{101}\!$ $I_{109}\!$ $K_{111}\!$ $L_{109}$ $M_{105}\!$
$N_{97}\!$ $O_{87}\!$ $Q_{77}\!$ $R_{65}\!$ $T_{53}\!$ $U_{43}$
$V_{33}\!$ $W_{24}\!$ $X_{18}\!$ $Y_{12}\!$ $Z_8\!$ $21_5\!$
$22_3\!$ $23$ $24)\ $ $^8(S_4\!$ $P_4\!$ $D_{12}\!$ $F_{11}\!$
$G_{17}\!$ $H_{15}\!$ $I_{19}\!$ $K_{16}\!$ $L_{18}\!$ $M_{14}\!$
$N_{14}\!$ $O_{10}\!$ $Q_{10}\!$ $R_6\!$
$T_6\!$ $U_3\!$ $V_3\!$ $W\!$ $X)\ $ $^{10}(PDFGHIKLMN)\ $ $^{12}S$\\
\end{longtable*}
\endgroup

\subsection{Derivation of the Even Rule for two equivalent electrons}

If only two equivalent electrons are involved, there is an ``Even
Rule'' \cite{ch63} which states
\begin{quote}
For two equivalent electrons the only states that are allowed are
those for which the sum $(L+S)$ is even.
\end{quote}

This rule can be deduced from our formulas as below. Based on
equations~(\ref{sub2}) and (\ref{sub1}), when $0\leq M_L \leq 2l$,
we have
\begin{equation}
A(2,l,l,0,M_L)
=\sum\limits_{M_{L_-}=\{-l,\;M_L-l\}_{\max}}^{\{l,\;M_L+l\}_{\min}}
1= {2l - M_L + 1}.
\end{equation}
Based on equations~(\ref{sub3}) and (\ref{sub1}), when $0 \leq M_L
\leq 2l-1$, we have
\begin{eqnarray}
A(2,l,l_b,2,M_L) &=& \sum_{M_{L_I} = \lfloor \frac{M_L}{2} + 1 \rfloor}^{\{l_b,\;M_L+l\}_{\min}}1 \label{even:7}\\
&=& l_b - \lfloor \frac{M_L}{2} \rfloor .
\end{eqnarray}

Hence, based on our main formula [equation~(\ref{main})], we have
\begin{eqnarray}
X(2,l,S',L) &=&\left\{ {\begin{array}{ll} {\lfloor \frac{L}{2}
\rfloor - \lfloor \frac{L-1}{2} \rfloor} & ~{\rm when}~
S'=0\\
{\lfloor \frac{L+1}{2} \rfloor -\lfloor \frac{L}{2} \rfloor} &
~{\rm when}~ S'=2 \end{array}}\right. \nonumber\\ &=&\lfloor
\frac{L+S}{2} \rfloor -\lfloor \frac{L+S-1}{2} \rfloor
.\label{even:12}
\end{eqnarray}
Therefore, only when $(L+S)$ is even, the function $X(N,l,S',L)$
is not vanish, viz. we get ``Even Rule''.

\subsection{Derivation of a new rule for three equivalent electrons}

Based on our theory [equations~(\ref{main})-(\ref{sub4})], we
derived a new rule for three equivalent electrons, which can be
stated as a formula below 
\begin{widetext}
\begin{equation}
\raggedright X(3,l,S',L)=\left\{ {\begin{array}{*{20}l} L-\lfloor
\frac{L}{3}
\rfloor & {\rm when}~S'=1,\, 0\leq L <l\\
l-\lfloor \frac{L}{3}
\rfloor & {\rm when}~S'=1,\, l\leq L \leq 3l-1\\
\lfloor \frac{L}{3}
\rfloor -\lfloor \frac{L-l}{2} \rfloor +\lfloor \frac{L-l+1}{2} \rfloor & {\rm when}~S'=3,\, 0\leq L <l\\
\lfloor \frac{L}{3} \rfloor -\lfloor \frac{L-l}{2} \rfloor & {\rm when}~S'=3,\, l\leq L \leq 3l-3\\
0 & {\rm other~cases} \\
\end{array}}\right. \label{tri:12}
\end{equation}
\end{widetext}

This rule can be derived respectively according to the two
possible values of S' ($S'=1$ or $3$).

\subsubsection{When $S'=1$}

To $S'=1$, we will derive the formula below
\begin{equation}
\raggedright X(3,l,S',L)=\left\{ {\begin{array}{*{20}l}L-\lfloor
\frac{L}{3}
\rfloor & when~S'=1,\, 0\leq L <l\\
l-\lfloor \frac{L}{3}
\rfloor & when~S'=1,\, l\leq L \leq 3l-1\\
0 & other~cases\\
\end{array}}\right.\label{tri:15}
\end{equation}
Based on equations~(\ref{sub1}), (\ref{sub2}), and~(\ref{even:7}),
when
\begin{equation*}
M_S'=1, \vert M_L \vert \leq
f(\frac{3-1}{2}-1)+f(\frac{3+1}{2}-1)=3l-1,
\end{equation*}
we have
\begin{widetext}
\begin{eqnarray}
\fl {A(3,l,l,1,M_L)
=~\hspace{-1em}\sum\limits_{M_{L_-}=\{-f(0),\;M_L-f(1)\}_{\max}}^{\{f(0),\;M_L+f(1)\}_{\min}}
\hspace{-1ex}\{A(1,l,l,1, M_{L_-}) A(2,l,l,2,M_L-~\!M_{L_-})\}}\nonumber\\
\fl{\hphantom{A(3,l,l,1,M_L)}=\sum\limits_{M_{L_-}=\{-l,\;M_L-2l+1\}_{\max}}^{\{l,\;M_L+2l-1\}_{\min}}
\{(l,M_L-M_{L_-}+l)_{\min}-\lfloor \frac{M_L-M_{L_-}}{2} \rfloor
\}}\nonumber\\
\fl{\hphantom{A(3,l,l,1,M_L)}=~\left\{ {
\begin{array}{*{20}l}
\sum\limits_{M_{L_-}={-l}}^l\{(l,M_L-M_{L_-}+l)_{\min}-\lfloor
\frac{M_L-M_{L_-}}{2} \rfloor
\} & {\raisebox{-1ex}[\height][0pt]{\framebox(18,16){:A}}}\\
\sum\limits_{M_{L_-}={M_L-2l+1}}^l\{l-\lfloor
\frac{M_L-M_{L_-}}{2}
\rfloor \} & {\raisebox{-1ex}[\height][0pt]{\framebox(18,16){:B}}}\\
\end{array}}\right.}\label{tri:16}
\end{eqnarray}
\end{widetext}
where ${\raisebox{-.5ex}[\height][0pt]{\framebox(18,16){:A}}}$
here means the case when $0 \leq M_L \leq l-1$, and
${\raisebox{-.5ex}[\height][0pt]{\framebox(18,16){:B}}}$ means the
case when $l-1 \leq M_L \leq 3l-1$.

Then based on equations~(\ref{sub1}), (\ref{sub3}),
and~(\ref{even:7}), when
\begin{equation*}
M_S'=3, \qquad \vert M_L \vert \leq f(2)=3l-3,
\end{equation*}
we have
\begin{widetext}
\begin{eqnarray}
\fl{A(3,l,l_b,3,M_L) =\sum\limits_{M_{L_I}=\lfloor \frac{M_L-1}{3}
+ \frac{3+1}{2} \rfloor}^{\{l_b,\;M_L+f(1)\}_{\min}} A(2,l,M_{L_I}-1,2,M_L - M_{L_I})}\nonumber\\
\fl{\hphantom{A(3,l,l_b,3,M_L)}=\hspace{-1.5ex}\sum\limits_{M_{L_I}
= \lfloor \frac{M_L-1}{3} + 2 \rfloor}^{l_b}
\hspace{-.75ex}\left\{(M_{L_I}-1,
M_L-M_{L_I}+l)_{\min}-\left\lfloor \frac{M_L-M_{L_I}}{2}
\right\rfloor \right\}}\label{tri:17}
\end{eqnarray}
\end{widetext}

Hence, when $S=1/2\;(S'=1),\ 0 \leq L \leq l-2$, we have
\begin{widetext}
\begin{eqnarray}
\fl{\Delta_1=A(3,l,l,1,L)-A(3,l,l,1,L+1)}\nonumber\\
\fl{\hphantom{\Delta_1}=\sum\limits_{M_{L_-}={-l}}^l\{(l,L-M_{L_-}+l)_{\min}-\lfloor
\frac{L-M_{L_-}}{2} \rfloor
\}}\nonumber\\
\fl{\hphantom{\Delta_1}\qquad-\sum\limits_{M_{L_-}={-l}}^l\{(l,L+1-M_{L_-}+l)_{\min}-\lfloor
\frac{L+1-M_{L_-}}{2} \rfloor
\}} \nonumber\\
\fl{\hphantom{\Delta_1}=\sum\limits_{M_{L_-}={-l}}^l\{\lfloor
\frac{L+1-M_{L_-}}{2} \rfloor -\lfloor \frac{L-M_{L_-}}{2}
\rfloor\}}\nonumber\\
\fl{\hphantom{\Delta_1}\qquad +(
\sum\limits_{M_{L_-}={-l}}^L+\sum\limits_{M_{L_-}={L+1}}^l)
\left\{(l,L-M_{L_-}+l)_{\min}-(l,L+1-M_{L_-}+l)_{\min}\right\}
}\nonumber\\
\fl{\hphantom{\Delta_1}= \left\lfloor {\frac{{L + l + 1}} {2}}
\right\rfloor  - \left\lfloor {\frac{{L - l}} {2}} \right\rfloor
\footnotemark[3]+0+\sum\limits_{M_{L_-}={L+1}}^l(-1)}\nonumber\\
\fl{\hphantom{\Delta_1}= \left\lfloor {\frac{{L + l + 1}} {2}}
\right\rfloor  - \left\lfloor {\frac{{L - l}} {2}}
\right\rfloor+(L-l)}
\end{eqnarray}
\end{widetext}
\footnotetext[3]{$\begin{array}{*2l} {\rm Use~the~formula~below~(a~and~b~are~integers)} \\
\qquad \sum\limits_{i=a}^b{\{\lfloor
\frac{i+1}{2}\rfloor-\lfloor\frac{i}{2}\rfloor\}}=\lfloor\frac{b+1}{2}\rfloor-\lfloor\frac{a}{2}\rfloor
\end{array}$}
and
\begin{widetext}
\begin{eqnarray}
\fl{\Delta_2=A(3,l,l,3,L)-A(3,l,l,3,L+1)}\nonumber\\
\fl{\hphantom{\Delta_2}=\sum\limits_{M_{L_I} = \lfloor
\frac{L-1}{3} + 2 \rfloor}^{l} \left\{(M_{L_I}-1,
L-M_{L_I}+l)_{\min}-\left\lfloor \frac{L-M_{L_I}}{2} \right\rfloor
\right\}} \nonumber\\
\fl{\hphantom{\Delta_2}\qquad-\sum\limits_{M_{L_I} = \lfloor
\frac{L}{3} + 2 \rfloor}^{l} \left\{(M_{L_I}-1,
L+1-M_{L_I}+l)_{\min}-\left\lfloor \frac{L+1-M_{L_I}}{2}
\right\rfloor \right\}}
 \nonumber\\
\fl{\hphantom{\Delta_2}=(\sum\limits_{M_{L_I} \leq \lfloor
\frac{L+l+1}{2} \rfloor}+\sum\limits_{M_{L_I} = \lfloor
\frac{L+l+1}{2} \rfloor +1}^{l}) \bigl\{(M_{L_I}-1,
L-M_{L_I}+l)_{\min}\bigr.}
\nonumber\\
\fl{\hphantom{\Delta_2}\qquad -\bigl. (M_{L_I}-1,
L+1-M_{L_I}+l)_{\min}\bigr\}} \nonumber\\
\fl{\hphantom{\Delta_2}\; +~\hspace{-1ex}\left\{
{\begin{array}{*{20}l} \hspace{-.5ex}\sum\limits_{M_{L_I} =
\lfloor \frac{L}{3}\rfloor + 2 }^l \left\{\left\lfloor
\frac{L+1-M_{L_I}}{2}
\right\rfloor -\left\lfloor \frac{L-M_{L_I}}{2} \right\rfloor \right\} & {\raisebox{-1ex}[\height][0pt]{\framebox(18,16){:A}}}\\
\hspace{-.5ex}\sum\limits_{M_{L_I} = \lfloor \frac{L}{3}\rfloor +
2}^l \hspace{-.5ex}\left\{\left\lfloor \frac{L+1-M_{L_I}}{2}
\right\rfloor \!-\!\left\lfloor \frac{L-M_{L_I}}{2} \right\rfloor
\right\}+\!(\lfloor \frac{L-1}{3}\rfloor +2)\!-\!1-\!\lfloor
\frac{L-(\lfloor \frac{L-1}{3}
\rfloor +2)}{2} \rfloor & {\raisebox{-1ex}[\height][0pt]{\framebox(18,16){:B}}} \\
\end{array}}\right.}\nonumber\\
\fl{\hphantom{\Delta_2}=0-\hspace{-1em}\sum\limits_{M_{L_I} =
\lfloor \frac{L+l+1}{2} \rfloor +1}^{l}\hspace{-1ex}1 + \left\{
{\begin{array}{*{20}l} \lfloor \frac{L - (\lfloor \frac{L}{3}
\rfloor +2)+1}{2}\rfloor - \lfloor \frac{L-l}{2}
\rfloor \footnotemark[3] & {\raisebox{-1ex}[\height][0pt]{\framebox(18,16){:A}}}\\
\lfloor \frac{L-(\lfloor \frac{L}{3} \rfloor+2)+1}{2}\rfloor -
\lfloor \frac{L-l}{2} \rfloor\footnotemark[3] -\left( \left\lfloor
\frac{L}{3}\right\rfloor-\left\lfloor \frac{L-(\lfloor
\frac{L}{3}\rfloor+1)}{2} \right\rfloor \right) & {\raisebox{-1ex}[\height][0pt]{\framebox(18,16){:B}}} \\
\end{array}}\right. } \nonumber\\
\fl{\hphantom{\Delta_2}= \left\lfloor \frac{L+l+1}{2}
\right\rfloor -l+\lfloor \frac{L}{3}\rfloor- \left\lfloor {\frac{L
- l}{2}} \right\rfloor }\label{tri:19}
\end{eqnarray}
\end{widetext}
where ${\raisebox{-.5ex}[\height][0pt]{\framebox(18,16){:A}}}$
here means the case when $\frac{L}{3}$ is not an integer, and
${\raisebox{-.5ex}[\height][0pt]{\framebox(18,16){:B}}}$ means the
case when $\frac{L}{3}$ is an integer. Thus we have
\begin{eqnarray}
\fl{X(3,l,1,L)=~\!A(3,l,l,1,L)-~\!A(3,l,l,1,L+~\!1)} \nonumber\\
\fl{\hphantom{X(3,l,1,L)}\qquad +~\!A(3,l,l,3,L+~\!1)-~\!A(3,l,l,3,L)}\nonumber \\
\fl{\hphantom{X(3,l,1,L)}=\Delta_1-\Delta_2} \nonumber \\
\fl{\hphantom{X(3,l,1,L)}= L -\left\lfloor {\frac{L}{3}}
\right\rfloor .}\label{tri:20-1}
\end{eqnarray}
When $L=l-1$, according to equation~(\ref{tri:16}),
$A(3,l,l,1,L+1)$ in equation~(\ref{tri:20-1}) equals
\begin{widetext}
\begin{eqnarray}
A(3,l,l,1,l)&=&\sum\limits_{M_{L_-}={-l+1}}^l\{l-\lfloor
\frac{l-M_{L_-}}{2} \rfloor
\}\nonumber\\
&=&\sum\limits_{M_{L_-}={-l}}^l\{l-\lfloor \frac{l-M_{L_-}}{2}
\rfloor \}-\left. \{l-\lfloor
\frac{l-M_{L_-}}{2} \rfloor \}\right|_{M_{L_-}=-l}\nonumber\\
&=&\sum\limits_{M_{L_-}={-l}}^l\{l-\lfloor \frac{l-M_{L_-}}{2}
\rfloor \},
\end{eqnarray}
\end{widetext}
which has the same value as in
equation~(\ref{tri:20-1}). Thus we can get the same expression of
function $\mathbf{X}$ also in this case.

When $S=1/2\;(S'=1),\ l \leq L \leq 3l-4$, we have
\begin{widetext}
\begin{eqnarray}
\fl{\Delta_1=A(3,l,l,1,L)-A(3,l,l,1,L+1)}\nonumber\\
\fl{\hphantom{\Delta_1}=\sum\limits_{M_{L_-}={L-2l+1}}^l \left(
l-\left\lfloor \frac{L-M_{L_-}}{2} \right\rfloor \right)
-\sum\limits_{M_{L_-}={L-2l+2}}^l \left( l-\left\lfloor
\frac{L+1-M_{L_-}}{2}
\right\rfloor \right)} \nonumber\\
\fl{\hphantom{\Delta_1}=\sum\limits_{M_{L_-}={L-2l+2}}^l \left\{
\left\lfloor \frac{L+1-M_{L_-}}{2} \right\rfloor -\left\lfloor
\frac{L-M_{L_-}}{2} \right\rfloor \right\} +\left( l-\left\lfloor
\frac{L-(L-2l+1)}{2}
\right\rfloor \right) }\nonumber\\
\fl{\hphantom{\Delta_1}= \left( \left\lfloor
\frac{L-(L-2l+2)+1}{2} \right\rfloor -\left\lfloor \frac{L-l}{2}
\right\rfloor
\right) +1}\nonumber\\
\fl{\hphantom{\Delta_1}= l - \left\lfloor {\frac{L - l}{2}}
\right\rfloor },
\end{eqnarray}
\end{widetext} and
\begin{widetext}
\begin{eqnarray}
\fl{\Delta_2=A(3,l,l,3,L)-A(3,l,l,3,L+1)}\nonumber\\
\fl{\hphantom{\Delta_2}=\hspace{-1.5ex}\sum\limits_{M_{L_I} =
\lfloor \frac{L-1}{3}\rfloor + 2 }^l
\left\{(M_{L_I}-1)-\left\lfloor \frac{L-M_{L_I}}{2} \right\rfloor
\right\}-\hspace{-1.5ex}\sum\limits_{M_{L_I} = \lfloor
\frac{L}{3}\rfloor + 2 }^l \left\{(M_{L_I}-1)-\left\lfloor
\frac{L+1-M_{L_I}}{2} \right\rfloor \right\}}
 \nonumber\\
\fl{\hphantom{\Delta_2}=~\hspace{-1ex}\left\{
{\begin{array}{*{20}l} \hspace{-.5ex}\sum\limits_{M_{L_I} =
\lfloor \frac{L}{3}\rfloor + 2 }^l \left\{\left\lfloor
\frac{L+1-M_{L_I}}{2}
\right\rfloor -\left\lfloor \frac{L-M_{L_I}}{2} \right\rfloor \right\} & {\raisebox{-1ex}[\height][0pt]{\framebox(18,16){:A}}}\\
\hspace{-.5ex}\sum\limits_{M_{L_I} = \lfloor \frac{L}{3}\rfloor +
2}^l \hspace{-.5ex}\left\{\left\lfloor \frac{L+1-M_{L_I}}{2}
\right\rfloor \!-\!\left\lfloor \frac{L-M_{L_I}}{2} \right\rfloor
\right\}+\!(\lfloor \frac{L-1}{3}\rfloor +2)\!-\!1-\!\lfloor
\frac{L-(\lfloor \frac{L-1}{3}
\rfloor +2)}{2} \rfloor & {\raisebox{-1ex}[\height][0pt]{\framebox(18,16){:B}}} \\
\end{array}}\right.}\nonumber\\
\fl{\hphantom{\Delta_2}=\left\{ {\begin{array}{*{20}l} \lfloor
\frac{L - ( \lfloor \frac{L}{3} \rfloor +2)+1}{2}\rfloor - \lfloor
\frac{L-l}{2}
\rfloor & {\raisebox{-1ex}[\height][0pt]{\framebox(18,16){:A}}}\\
 \lfloor \frac{L-(\lfloor \frac{L}{3}
\rfloor+2)+1}{2}\rfloor - \lfloor \frac{L-l}{2} \rfloor -\left(
\left\lfloor \frac{L}{3}\right\rfloor-\left\lfloor
\frac{L-(\lfloor
\frac{L}{3}\rfloor+1)}{2} \right\rfloor \right) & {\raisebox{-1ex}[\height][0pt]{\framebox(18,16){:B}}} \\
\end{array}}\right.} \nonumber\\
\fl{\hphantom{\Delta_2}= \left\lfloor \frac{L}{3}\right\rfloor -
\left\lfloor {\frac{L - l}{2}} \right\rfloor }\label{tri:23}
\end{eqnarray}
\end{widetext}
where ${\raisebox{-.5ex}[\height][0pt]{\framebox(18,16){:A}}}$
here means the case when $\frac{L}{3}$ is not an integer, and
${\raisebox{-.5ex}[\height][0pt]{\framebox(18,16){:B}}}$ means the
case when $\frac{L}{3}$ is an integer. Thus we have
\begin{eqnarray}
\fl{X(3,l,1,L)=~\!A(3,l,l,1,L)-~\!A(3,l,l,1,L+~\!1)}\nonumber\\
\fl{\hphantom{X(3,l,1,L)}\qquad +~\!A(3,l,l,3,L+~\!1)-~\!A(3,l,l,3,L)}\nonumber \\
\fl{\hphantom{X(3,l,1,L)}=\Delta_1-\Delta_2} \nonumber \\
\fl{\hphantom{X(3,l,1,L)}= l -\left\lfloor {\frac{L}{3}}
\right\rfloor .}\label{tri:20}
\end{eqnarray}
When $L=3l-3$, $A(3,l,l,3,L+1)$ vanishes; when $L=3l-2$,
$A(3,l,l,3,L)$ also vanishes; when $L=3l-1$, $A(3,l,l,1,L+1)$ also
vanishes; and we can get the function $\mathbf{X}$ which equals
$1$, coinciding with equation~(\ref{tri:20}).

Combining equations~(\ref{tri:20-1}) and (\ref{tri:20}), we get
the equation~(\ref{tri:15}).

\subsubsection{When $S'=3$}

To $S'=3$, we will derive the formula below
\begin{widetext}
\begin{equation}
\raggedright X(3,l,S',L)=\left\{ {\begin{array}{*{20}l}
\lfloor
\frac{L}{3}
\rfloor -\lfloor \frac{L-l}{2} \rfloor +\lfloor \frac{L-l+1}{2} \rfloor & when~S'=3,\, 0\leq L <l-1\\
\lfloor \frac{L}{3}
\rfloor -\lfloor \frac{L-l}{2} \rfloor & when~S'=3,\, l-1\leq L \leq 3l-3\\
0 & other~cases\\
\end{array}}\right.\label{tri:25}
\end{equation}
\end{widetext}
Based on equations~(\ref{tri:19}), when
\begin{equation*}
S=3/2\;(S'=3),\ 0 \leq L \leq l-1,
\end{equation*}
we have
\begin{eqnarray}
\Delta_1 &=& A(3,l,l,3,L)-A(3,l,l,3,L+1) \nonumber\\
&=&\left\lfloor \frac{L-l+1}{2} \right\rfloor +\lfloor
\frac{L}{3}\rfloor- \left\lfloor {\frac{L - l}{2}} \right\rfloor
\end{eqnarray}
$\Delta_2$ just vanishes, thus we have
\begin{eqnarray}
\fl{X(3,l,3,L)=~\!\!A(3,l,l,3,L)-~\!\!A(3,l,l,3,L+~\!1)}\nonumber \\
\fl{\hphantom{X(3,l,3,L)}\qquad +~\!\!A(3,l,l,5,L+~\!1)-~\!\!A(3,l,l,5,L)}\nonumber \\
\fl{\hphantom{X(3,l,3,L)}=\Delta_1-\Delta_2} \nonumber \\
\fl{\hphantom{X(3,l,3,L)}=\lfloor \frac{L}{3}\rfloor- \left\lfloor
{\frac{L - l}{2}} \right\rfloor +\left\lfloor \frac{L-l+1}{2}
\right\rfloor}\label{tri:27}
\end{eqnarray}

Based on equations~(\ref{tri:23}), when
\begin{equation*}
S=3/2\;(S'=3),\ l \leq L \leq 3l-4,
\end{equation*}
we have
\begin{equation}
\Delta_1 = A(3,l,l,3,L)-A(3,l,l,3,L+1)=\left\lfloor
\frac{L}{3}\right\rfloor - \left\lfloor {\frac{L - l}{2}}
\right\rfloor\label{tri:28}
\end{equation}
When $L=3l-3$, $A(3,l,l,3,L+1)$ vanishes, and we can get the
function $\Delta_1$ equaling $1$, which also can be expressed by
equation~(\ref{tri:28}). $\Delta_2$ also vanishes, thus we have
\begin{eqnarray}
\fl{X(3,l,3,L)=~\!\!A(3,l,l,3,L)-~\!\!A(3,l,l,3,L+~\!1)}\nonumber\\
\fl{\hphantom{X(3,l,3,L)}\qquad +~\!\!A(3,l,l,5,L+~\!1)-~\!\!A(3,l,l,5,L)}\nonumber \\
\fl{\hphantom{X(3,l,3,L)}=\Delta_1-\Delta_2} \nonumber \\
\fl{\hphantom{X(3,l,3,L)}=\lfloor \frac{L}{3}\rfloor- \left\lfloor
{\frac{L - l}{2}} \right\rfloor}\label{tri:29}
\end{eqnarray}
Combining equations~(\ref{tri:27}) and (\ref{tri:29}), we get the
equation~(\ref{tri:25}). Banding together both of
equations~(\ref{tri:15}) and (\ref{tri:25}), we naturally get the
rule for three equivalent electrons [equation~(\ref{tri:12})].

\section{CONCLUSION}

Mainly based on a digital counting procedure, the alternative
mathematical technique to determine the LS spectral terms arising
from $l^N$ configurations, is immediately applicable for studies
involving one orbital shell model. It makes the calculation of
coupled states of excited high energy electrons possible, and
offered a basis for the further calculations of energy levels for
laser and soft X-ray. Though the derivation of our theory is a
little complicated and thus is presented in Appendix below.
Compared to other theoretical methods reported earlier in
literature, this method is much more compact, and especially
offered a direct way in calculation.

In addition, based on this alternative mathematical basis, we may
also try to calculate the statistical distribution of J-values for
$l^N$ configurations~\cite{bauche}, and try to deduce some more
powerful rules or formulas probably could be deduced for
determining the LS terms, such as equations~(\ref{even:12}) and
(\ref{tri:12}). Indeed, it may also be applicable to other
coupling schemes.

\appendix


\section{Derivation of the main formula equation~(\ref{main})}

Now we'll determine the number of spectral terms having total
orbital angular quantum number $L$ and total spin quantum number
$S'/2$ arising from $l^N$ electronic configurations, which is
denoted by $X(N,l,S',L)$.

The number of spectral terms having allowed orbital magnetic
quantum number $L_0$ and spin magnetic quantum number $S_0'/2$ in
$l^N$ electronic configurations equals $A(N,l,l, S_0', L_0)$,
namely the number of spectral terms with $L \geq L_0, \;S \geq
S_0'/2$. And these spectral terms can also be subdivided according
to their quantum numbers of $L$ and $S$ into four types as
follows:

\textcircled{\small 1}. $L=L_0, S= S_0'/2$: the number of this
type is $X(N,l,S_0',L_0)$.

\textcircled{\small 2}. $L=L_0, S \geq \frac{S_0'}{2}+1$: the
number of this type equals
\begin{equation}
A(N,l,l,S_0'+2,L_0)-A(N,l,l,S_0'+2, L_0+1).\nonumber
\end{equation}

\textcircled{\small 3}. $L \geq L_0+1, S= S_0'/2$: the number of
this type equals
\begin{equation}
A(N,l,l,S_0',L_0+1)-A(N,l,l,S_0'+2,L_0+1).\nonumber
\end{equation}

\textcircled{\small 4}. $L \geq L_0+1, S \geq \frac{S_0'}{2} + 1$:
the number of this type is $A(N,l,l,S_0'+2,L_0 +1)$.

Hence, we have
\begin{eqnarray}
\fl{A(N,l,l,S_0',L_0)=X(N,l,S_0',L_0)+A(N,l,l,S_0'+2,L_0+1)}\nonumber\\
\fl{\hphantom{A(N,l,l,S_0',L_0
)=}+\left\{A(N,l,l,S_0'+2,L_0)-A(N,l,l,S_0'+2,L_0+1)\right\}}
\nonumber\\
\fl{\hphantom{A(N,l,l,S_0',L_0
)=}+\left\{A(N,l,l,S_0',L_0+1)-A(N,l,l,S_0'+2,L_0+1)\right\}.}
\end{eqnarray}
Therefore
\begin{eqnarray}
\fl{X(N,l,S',L)=A(N,l,l,S',L)+A(N, l, l, S'+2,L+1)-A(N, l, l, S'+2, L)}\nonumber\\
\fl{\hphantom{X(N,l,S',L)=}\qquad-A(N, l, l, S',L+1).}
\end{eqnarray}

\section{Derivation of equation~(\ref{sub1})}

For one-particle configurations $(N=1)$, there is only one
spectral term. Thus toward any allowable value of $M_L$, we have
\begin{equation}
A(1,l,l_b,1,M_L)=1 \qquad (-l\leq M_L\leq l).
\end{equation}

\section{Derivation of equation~(\ref{sub2})}

In this case $(N\geq 2)$, there are some electrons spin-up and
others spin-down. Taking account of the Pauli principle, we sort
the N electrons into two classes: (1) Spin-down electrons
class~\textcircled{-} consists of $k_-~(\geq 1)$ electrons with
$m_{s_i}=-1/2$ $(i=1,2,\ldots,k_-)$; (2) Spin-up electrons
class~\textcircled{+} consists of $k_+~(\geq 1)$ electrons with
$m_{s_j}=1/2$ $(j=1,2,\ldots,k_+)$. In each class, the orbital
magnetic quantum number of each electron is different from each
other. The total spin and orbital magnetic quantum number for
class \textcircled{-} are
\begin{equation}
M_{s_-}=\sum\limits_{i=1}^{k_-}m_{s_i}=-\frac{k_-}{2} \qquad
M_{L_-}=\sum\limits_{i=1}^{k_-}m_{l_i}.
\end{equation}
For class \textcircled{+},
\begin{equation}
M_{s_+}'=2M_{s_+}=2 \sum\limits_{j=1}^{k_+}m_{s_j}=k_+ \qquad
M_{L_+}= \sum\limits_{j=1}^{k_+}m_{l_j}.
\end{equation}

\subsection{The number of permitted states to each $M_{L_-}$
value}

When $M_L$ is fixed, for each allowable value of $M_{L_-}$, there
is a unique corresponding value of $M_{L_+}=M_L-M_{L_-}$. We can
denote by $A(k_-,l,l,M_{s_-}',M_{L_-})$ the number of permitted
states of the $k_-$ electrons in class \textcircled{-} according
to the notations defined in Section II. Based on any LS term
having a spin magnetic quantum number $M_S$ must also have a spin
magnetic quantum number $-M_S$, we have
\begin{eqnarray}
A(k_-,l,l,M_{s_-}',M_{l_-})&=&A(k_-,l,l,-M_{s_-}',M_{l_-})\nonumber\\
&=&A(k_-,l,l,k_-,M_{l_-}).
\end{eqnarray}
Correspondingly we denote by
$A(k_+,l,l,M_{s_+}',M_{L_+})=A(k_+,l,l,k_+,M_L-M_{L_-})$ for class
\textcircled{+}. Hence, to any value of $M_{L_-}$, the total
number of permitted states of $l^N$ is
$A(k_-,l,l,k_-,M_{L_-})\;A(k_+,l,l,k_+,M_L-M_{L_-})$.

\subsection{Determination of the range of $M_{L_-}$}

Firstly, the value of $\sum\limits_{i=1}^{k_-}m_{l_i}$ is minimum,
when the orbital magnetic quantum numbers of the $k_-$ electrons
in class \textcircled{-} respectively are
$-l,-(l-1),\ldots,-(l-k_-+1)$. Thus we have
\begin{equation}
(M_{L_-})_{\min} \geq
(\sum\limits_{i=1}^{k_-}{m_{l_i}})_{\min}=-\sum_{m=0}^{k_--1}(l -
m). \label{eq:9}
\end{equation}
Similarly, the value of $\sum\limits_{j=1}^{k_+}m_{l_j}$ is
maximum, when the orbital magnetic quantum numbers of the $k_+$
electrons in class \textcircled{+} respectively are
$l,(l-1),\ldots,(l-k_++1)$. Thus we have
\begin{equation}
(M_{L_-})_{\min} \geq
M_L-(M_{L_+})_{\max}=M_L-\sum_{m=0}^{k_+-1}(l-m). \label{eq:10}
\end{equation}
Comparing the equation~(\ref{eq:9}) with equation~(\ref{eq:10}),
we have
\begin{equation}
(M_{L_-})_{\min}=\left\{-\sum\limits_{m=0}^{k_--1}(l-m),\;M_L-\sum_{m=0}^{k_+-1}(l-m)\right\}_{\max}
\end{equation}

Similarly, due to
\begin{eqnarray}
(M_{L_-})_{\max} \!&\leq &\!
(\sum\limits_{i=1}^{k_-}m_{l_i})_{\max}=\sum\limits_{m=0}^{k_--1}(l - m),\\
(M_{L_-})_{\max} \!&\leq &\! M_L-(M_{L_+})_{\min}=
M_L+\sum\limits_{m=0}^{k_+-1}(l-m),\qquad
\end{eqnarray}
we have
\begin{equation}
(M_{L_-})_{\max}=\left\{\sum\limits_{m=0}^{k_--1}(l-m),\;M_L+\sum\limits_{m=0}^{k_+-1}(l-m)\right\}_{\min}
\end{equation}

\subsection{The total number of permitted states}

Recalling the relationship among $k_+$, $k_-$ and $M_S'$, $N$,
\begin{numcases}{}
N=k_++k_-,\\
M_S'=2M_S=2M_{s_-}+2M_{s_+}=k_+-k_-,
\end{numcases}
we have
\begin{equation}
k_-=(N-M_S')/2 \qquad k_+=(N+M_S')/2.
\end{equation}
Consequently, we get
\begin{widetext}
\begin{eqnarray}
\fl{A(N,l,l,M_S',M_L)=\hspace{-1em}\sum\limits_{M_{L_-}=(M_{L_-})_{\min}}^{(M_{L_-})_{\max}}\hspace{-1ex}\{A(k_-,l,l,k_-,M_{L_-})\;A(k_+,l,l,k_+,M_L-M_{L_-})\}}
\nonumber\\
\fl{\hphantom{A(N,l,l,M_S',M_L)}=\hspace{-1.25em}\sum\limits_{M_{L_-}=\{-f(\frac{N-M_S'}{2}-1),\;
M_L-f(\frac{N+M_S'}{2}-1)\}_{\max}}^{\{f(\frac{N-M_S'}{2}-1),\;M_L+f(\frac{N+M_S'}{2}-1)\}_{\min}}
\hspace{-1.5em}\left\{A(\frac{N-M_S'}{2},l,l,\frac{N-M_S'}{2},M_{L_-})\right.}\nonumber\\
\fl{\hphantom{A(N,l,l,M_S',M_L)}\qquad\left.\times
A(\frac{N+M_S'}{2},l,l,\frac{N+M_S'}{2},M_L-M_{L_-})\right\}.}
\end{eqnarray}
\end{widetext}

\subsection{The domain of definition}

\subsubsection{The range of $M_L$ and $L$}

Based on
\begin{equation}
(M_{L_-})_{\min}+(M_{L_+})_{\min} \leqslant M_L \leqslant
(M_{L_-})_{\max}+(M_{L_+})_{\max}, \\
\end{equation}
and
\begin{equation}
(M_{L_-^+})_{\min}=-\sum\limits_{m=0}^{k_-^+-1}{(l-m)},\
(M_{L_-^+})_{\max}=\sum\limits_{m=0}^{k_-^+-1}{(l-m)},
\end{equation}
we have
\begin{equation}
\vert M_L \vert \leqslant
f(\frac{N-M_S'}{2}-1)+f(\frac{N+M_S'}{2}-1).
\end{equation}

Therefore, the total orbital angular quantum number $L$ must
fulfil the inequality
\begin{equation}
0 \leqslant L \leqslant
f(\frac{N-M_S'}{2}-1)+f(\frac{N+M_S'}{2}-1).\label{range:L}
\end{equation}

\subsubsection{The range of $M_S'$ and $S'$}

Concerning
\begin{eqnarray}
M_S'=k_+-k_-=k_+-(N-k_+)=2k_+ - N,\\
(k_+)_{\min}=1 \qquad (k_+)_{\max}=N-1,
\end{eqnarray}
we have
\begin{equation}
\{M_S'\}=\{2-N,4-N,\ldots,N-4,N-2\}.
\end{equation}
and
\begin{eqnarray}
\{S'\}= \left\{{\begin{array}{*{2}l} \{0,2,\ldots,N-2\}
\qquad & ({\rm N \quad even})\\
\{1,3,\ldots,N-2\} \qquad & ({\rm N \quad odd})
\end{array}} \right.
\end{eqnarray}
Now we reduce the two expressions into one expression
\begin{equation}
\{S'\}=\{N-2\lfloor N/2 \rfloor,N+2-2\lfloor N/2
\rfloor,\ldots,N-2\}.\label{range:S'}
\end{equation}

Therefore, the equation~(\ref{sub2}) has been proved completely.

\section{Derivation of equation~(\ref{sub3})}

Now we discuss the case that all of the $N$ electrons are spin-up,
namely
\begin{equation}
M_S'=N\quad and\quad M_S=N/2.\label{range:S2}
\end{equation}
Based on the Pauli exclusion principle, we can prescribe
\begin{equation}
m_{l_1} > m_{l_2} > \ldots > m_{l_i} > \ldots > m_{l_N}.
\label{ieq:29}
\end{equation}
Now we also treat these electrons as two classes: (1) Class
\textcircled{\tiny I} consists of the electron whose orbital
magnetic quantum number is largest; (2) Class \textcircled{\tiny
II} consists of the other electrons. Thus the total orbital
magnetic quantum number of the two classes are
\begin{equation}
M_{L_I} = m_{l_1} \qquad M_{L_{II}} = \sum\limits_{i = 2}^N
{m_{l_i}}
\end{equation}

\subsection{The number of permitted states to each
$M_{L_I}$ value}

In view of inequality~(\ref{ieq:29}), we have
\begin{equation}
(m_{l_{i + 1}})_{\max} = m_{l_i} - 1\qquad
\quad(i=1,2,\ldots,N-1).\label{rangeD:max}
\end{equation}
Thus we can denote by $A(N-1,l,l_b,N-1,M_{L_{II}})$ which equals
$A(N-1,l,M_{L_I}-1,N-1,M_L -M_{L_I})$, the permitted states of
class \textcircled{\tiny II} consisting of the latter $(N-1)$
electrons, according to the notations prescribed in Section II.
For any allowed value of $M_{L_I}$, there is only one state for
class \textcircled{\tiny I}. Therefore, to each value of
$M_{L_I}$, the total number of permitted states of the $N$
electrons is $A(N-1,l,M_{L_I} - 1,N-1,M_L -M_{L_I})$.

Then in class \textcircled{\tiny II}, we can treat $m_{l_2}$ as
$M_{L_I}$, and the latter $(\sum\limits_{i = 3}^N {m_{l_i}})$ as
$M_{L_{II}},\ldots$. Just continue our operation in this way,
after $(N-1)$ times of operation, and then based on
equation~(\ref{sub1}), we can get the final value of
$A(N,l,l_b,N,M_L)$.

\subsection{The range of $M_{L_I}$}

Based on
\begin{eqnarray}
(M_{L_I})_{\max} &\leqslant & M_L - (M_{L_{II}})_{\min} = M_L +
\sum\limits_{m = 0}^{N - 2} {(l - m)}, \qquad\\
(M_{L_I})_{\max} &\leqslant & l_b,
\end{eqnarray}
we have
\begin{equation}
(M_{L_I})_{\max} = \{l_b,\;M_L + \sum\limits_{m = 0}^{N - 2} {(l -
m)} \}_{\min}.
\end{equation}

In the following, we will prove
\begin{equation}
(M_{L_I})_{\min} = \lfloor \frac{M_L - 1}{N} + \frac{N + 1}{2}
\rfloor.
\end{equation}

Because of the symmetrical situation between $M_L>0$ and $M_L<0$
to a certain LS term, it is necessary only to consider the part of
the case which corresponds to $M_L \geq 0$.

In the case of $M_{L_I}$ being minimum, we have
\begin{equation}
(m_{l_2})_{\min}=M_{L_I}-2,\ (m_{l_{i + 1}})_{\min} = m_{l_i} - 1,
\label{rangeD:min}
\end{equation}
where (i=2,\ldots,N-1). Based on equations~(\ref{rangeD:max}) and
(\ref{rangeD:min}), we get the maximum value of $M_L$
\begin{eqnarray}
(M_L)_{\max} &=& (M_{L_I})_{\min} + (M_{L_{II}})_{\max}\nonumber\\
&=& (M_{L_I} )_{\min } + \sum\limits_{i = 2}^N \{(M_{L_I})_{\min }
- (i - 1)\}\nonumber\\ &=& N(M_{L_I})_{\min} - \frac{N(N-1)}{2};
\label{C37}
\end{eqnarray}
and the minimum value of $M_L$
\begin{eqnarray}
(M_L)_{\min} &=& (M_{L_I})_{\min} + (M_{L_{II}})_{\min}
\nonumber\\ &=&
(M_{L_I})_{\min} + \sum\limits_{i=2}^N \{(M_{L_I})_{\min}-i\} \nonumber\\
&=& N(M_{L_I})_{\min}-\frac{N(N-1)}{2} - (N - 1).\qquad
\label{C38}
\end{eqnarray}

Therefore
\begin{eqnarray}
M_L = N(M_{L_I})_{\min} - \frac{N(N - 1)}{2}-j,\\
(M_{L_I})_{\min} = \frac{M_L + j}{N} + \frac{N - 1}{2},
\end{eqnarray}
where $j$ could be $0,1,\ldots,$ and $N-1$, which just to make
sure that $(M_{L_I})_{\min}$ is an integer. Thus
\begin{equation}
(M_{L_I})_{\min} = \lfloor \frac{M_L + N - 1}{N} + \frac{N - 1}{2}
\rfloor = \lfloor \frac{M_L - 1}{N} + \frac{N + 1}{2} \rfloor.
\end{equation}

Consequently, we get
\begin{widetext}
\begin{eqnarray}
\fl{A(N,l,l_b,N,M_L) = \sum\limits_{M_{L_I} =
(M_{L_I})_{\min}}^{(M_{L_I})_{\max}} {A(N-1,l, M_{L_I} - 1,N - 1,M_L - M_{L_I})}}\nonumber \\
\fl{\hphantom{A(N,l,l_b,N,M_L)}=\hspace{-1em} \sum\limits_{M_{L_I}
= \lfloor \frac{M_L - 1}{N} + \frac{N+1}{2} \rfloor}^{\{l_b,\;M_L
+ \sum\limits_{m = 0}^{N - 2} {(l - m)} \}_{\min}}\hspace{-2em}
{A(N - 1,l,M_{L_I} - 1,N - 1,M_L - M_{L_I})}}
\end{eqnarray}
\end{widetext}

\subsection{The range of $M_L$ and $L$}

Based on
\begin{equation}
(M_L)_{\min}= - \sum\limits_{m = 0}^{N - 1} {(l - m)} \qquad
(M_L)_{\max}=\sum\limits_{m = 0}^{N - 1} {(l - m)},
\end{equation}
we have
\begin{equation}
\vert M_L \vert \leqslant f(N - 1).
\end{equation}

Hence the total orbital angular quantum number $L$ must fulfil
\begin{equation}
0 \leqslant L \leqslant f(N - 1).
\end{equation}

So, the equation~(\ref{sub3}) has been proved completely.

Now we have completely proved the five formulas represented in
Section II.

\begin{acknowledgments}
The authors are grateful to Prof. Jacques Bauche (Universit\'{e}
PARIS XI) for many useful discussions.
\end{acknowledgments}



\end{document}